\documentclass[ChapterTOCs,krantz1]{krantz}

\usepackage{amssymb,amsmath,bbm,bm}
\usepackage{graphicx,subfigure}
\usepackage{makeidx}
\usepackage{multicol}
\usepackage{epsfig,color}
\usepackage{wasysym}

\usepackage[sectionbib]{bibunits}
\usepackage{liealg}

\newcommand{\be}{\begin{eqnarray}}
\newcommand{\ee}{\end{eqnarray}}
\newcommand{\bea}{\begin{eqnarray}}
\newcommand{\eea}{\end{eqnarray}}
\newcommand{\bma}{\begin{subequations}}
\newcommand{\ema}{\end{subequations}}

\newcommand{\omicron}[1]{\textrm{O} \left[ \frac{1}{#1} \right] }

\def\bbbc{{\mathchoice {\setbox0=\hbox{$\displaystyle\rm C$}\hbox{\hbox
to0pt{\kern0.4\wd0\vrule height0.9\ht0\hss}\box0}}
{\setbox0=\hbox{$\textstyle\rm C$}\hbox{\hbox
to0pt{\kern0.4\wd0\vrule height0.9\ht0\hss}\box0}}
{\setbox0=\hbox{$\scriptstyle\rm C$}\hbox{\hbox
to0pt{\kern0.4\wd0\vrule height0.9\ht0\hss}\box0}}
{\setbox0=\hbox{$\scriptscriptstyle\rm C$}\hbox{\hbox
to0pt{\kern0.4\wd0\vrule height0.9\ht0\hss}\box0}}}}

\renewcommand{\theequation}{\thesection.\arabic{equation}}
\newcommand{\beq}{\begin{eqnarray}}
\newcommand{\eeq}{\end{eqnarray}}

\def\simge{\mathrel{%
       \rlap{\raise 0.511ex \hbox{$>$}}{\lower 0.511ex \hbox{$\sim$}}}}
\def\simle{\mathrel{
       \rlap{\raise 0.511ex \hbox{$<$}}{\lower 0.511ex \hbox{$\sim$}}}}

\newcommand{\BQ}{\begin{equation}}
\newcommand{\EQ}{\end{equation}}
\newcommand{\BQA}{\begin{eqnarray}}
\newcommand{\EQA}{\end{eqnarray}}

\def\stopenone{\leavevmode\hbox{\small1\kern-4.2pt\normalsize1}}



\makeindex

\begin{document}
\bibliographyunit[\chapter]
\defaultbibliography{biblio}
\defaultbibliographystyle{plain}
\title{Quantum phase transitions in two-dimensional electron systems}
\author{Alexander Shashkin, Institute of Solid State Physics, Chernogolovka, Moscow District 142432, Russia\\
Sergey Kravchenko, Physics Department, Northeastern University, Boston, Massachusetts 02115, U.S.A.}

\maketitle

%\frontmatter
%\include{frontmatter/Foreword}
%\include{frontmatter/preface}
%\include{frontmatter/introduction}

%\listoffigures
%\listoftables
\tableofcontents

%\mainmatter

%\include{Parts/symbollist}

\setcounter{page}{1}

%\part{Experimental Realizations of Quantum Phases and Quantum Phase Transitions}
\section{Strongly and Weakly Interacting 2D Electron Systems}

Two-dimensional (2D) electron systems are realized when the electrons
are free to move in a plane but their motion perpendicular to the
plane is quantized in a confining potential well. Quantum phase transitions realized experimentally in such
systems so far include metal-insulator transitions in
perpendicular magnetic fields, metal-insulator transition in zero
magnetic field, and possible transition to a Wigner crystal. The first
transition is governed by the externally controlled electron density
or magnetic field, while the other two are governed by the electron density. At low electron
densities in 2D systems, the strongly-interacting limit is reached
because the kinetic energy is overwhelmed by the energy of
electron-electron interactions. The interaction strength is
characterized by the ratio between the Coulomb energy and the Fermi
energy, $r_s^*=E_{ee}/E_F$. Assuming that the effective electron mass
is equal to the band mass, the interaction parameter $r_s^*$ in the
single-valley  case reduces to the Wigner-Seitz radius, $r_s=1/(\pi
n_s)^{1/2}a_B$, and therefore increases as the electron density,
$n_s$, decreases (here $a_B$ is the Bohr radius in the semiconductor).
Possible candidates for the ground state of the system include a Wigner
crystal characterized by spatial and spin ordering~\cite{SK:wigner34},
a ferromagnetic Fermi liquid with spontaneous spin ordering
\cite{SK:stoner46}, a paramagnetic Fermi liquid~\cite{SK:landau57}, etc. In
the strongly-interacting limit ($r_s\gg1$), no analytical theory has
been developed to date. According to numerical simulations
\cite{SK:tanatar89}, Wigner crystallization is expected in a very dilute
regime, when $r_s$ reaches approximately 35.  Refined numerical simulations~\cite{SK:attaccalite02} have predicted that prior to the
crystallization, in the range of the interaction parameter $25\leq
r_s\leq35$, the ground state of the system is a strongly correlated
ferromagnetic Fermi liquid. At higher electron densities, $r_s\sim1$,
the electron liquid is expected to be paramagnetic, with the
effective mass, $m$, and Land\'e $g$ factor renormalized by
interactions. Apart from the ferromagnetic Fermi liquid, other
intermediate phases between the Wigner crystal and the paramagnetic
Fermi liquid may also exist.

In real 2D electron systems, the inherent disorder leads to a drastic
change of the above picture, which significantly complicates the
problem. According to the scaling theory of localization
\cite{SK:abrahams79}, all electrons in a disordered infinite
noninteracting 2D system become localized at zero temperature and
zero magnetic field. At finite temperatures, regimes of strong and
weak localizations are distinguished: (i) if the conductivity of the
2D electron layer is activated, the resistivity diverges
exponentially as $T\rightarrow0$; and (ii) in the opposite limit of
weak localization the resistivity increases logarithmically with
decreasing temperature, an effect originating from the increased
probability of electron backscattering from impurities to the
starting point. Interestingly, the incorporation of weak interactions
($r_s<1$) between the electrons promotes the localization
\cite{SK:altshuler80}. However, for weak disorder and $r_s\geq1$ a
possible metallic ground state was predicted~\cite{SK:finkelstein83}.

In view of the competition between the interactions and disorder,
high- and low-disorder limits can be considered. In highly-disordered
electron systems, the range of low densities is not accessible as the
strong (Anderson) localization sets in. This corresponds to the
weakly-interacting limit in which an insulating ground state is
expected. The case of low-disordered electron systems is much more
interesting because low electron densities corresponding to the
strongly-interacting limit become accessible. According to the
renormalization group analysis for multi-valley 2D systems~\cite{SK:punnoose05}, strong electron-electron interactions can
stabilize the metallic ground state, leading to the existence of a
metal-insulator transition in zero magnetic field.

In quantizing magnetic fields, the interaction strength is
characterized by the ratio between the Coulomb energy and the
cyclotron splitting. In the ultra-quantum limit, it is similar to the
interaction parameter $r_s^*$. Within the concept of single-parameter
scaling for noninteracting 2D electrons~\cite{SK:pruisken88}, there is
only one extended state in the Landau level, and the localization
length diverges at the center of the Landau level~\cite{SK:iordansky82}.
For consistency with the scaling theory of localization in zero
magnetic field, it was predicted that extended states in the
Landau levels cannot disappear discontinuously with decreasing
magnetic field but must ``float up'' (move up in energy) indefinitely in the limit~\cite{SK:khmelnitskii84} of $B\rightarrow0$. The corresponding phase
diagram plotted in disorder versus inverse filling factor
($1/\nu=eB/hcn_s$) plane is known as the global phase diagram for the
quantum Hall effect (QHE)~\cite{SK:kivelson92}.  As long as no merging of the
extended states was considered to occur, their piercing of the Fermi
level was predicted to cause quantization of the Hall conductivity in
weak magnetic fields~\cite{SK:khmelnitskii92}. The case of strongly
interacting 2D electrons in the quantum Hall regime has not been
considered theoretically. In the very dilute regime, there are
theoretical predictions that Wigner crystallization is promoted in
the presence of a magnetic field (see, e.g., Ref.~\cite{SK:lozovik75}).

In this chapter, attention is focused on experimental results
obtained in low-disordered strongly interacting 2D electron systems,
in particular, (100)-silicon metal-oxide-semiconductor field-effect
transistors (MOSFETs). Due to the relatively large effective mass,
relatively small dielectric constant, and the presence of two valleys
in the spectrum, the interaction parameter in silicon MOSFETs is an
order of magnitude bigger at the same electron density than in
the 2D electron system in GaAs/AlGaAs heterostructures. Except at
very low electron densities, the latter electron system can be
considered weakly interacting. It is worth noting that the observed
effects of strong electron-electron interactions are more pronounced
in silicon MOSFETs compared to GaAs/AlGaAs heterostructures, although
the fractional QHE, which is usually attributed to
electron-electron interactions, has not been reliably established in
silicon MOSFETs.

\section{Proof of the Existence of Extended States in the Landau Levels}
\label{SK:proof}

In a magnetically quantized 2D electron system, the Landau levels 
bend up at the sample edges due to the confining potential, and edge
channels are formed where these intersect the Fermi energy (see,
e.g., Ref.~\cite{SK:halperin82}). There arises a natural question as to whether
the current in the quantum Hall state\footnote{In this state the Hall resistivity,
$\rho_{xy}=h/\nu e^2$, is quantized at integer filling factor $\nu$,
accompanied by vanishing longitudinal resistivity, $\rho_{xx}$
\cite{SK:klitzing80}.} flows in the bulk or at the edges of the sample.
Although the Hall conductivity $\sigma_{xy}$ was not directly
measured in early experiments on the QHE, it seemed
obvious that this value corresponds to the Hall resistivity
$\rho_{xy}$, in agreement with the concept of currents that flow in
the bulk~\cite{SK:QHE87}; it stands to reason that finite
$\sigma_{xy}$ would give evidence for the existence of extended
states in the Landau levels~\cite{SK:halperin82,SK:levine83}. This concept
was challenged by the edge current model~\cite{SK:buttiker88}. In the
latter approach extended states in the bulk are not crucial and the
problem of current distributions in the QHE is
reduced to a one-dimensional task in terms of transmission and
reflection coefficients as defined by the backscattering current at
the Fermi level between the edges. Importantly, if the edge current
contributes significantly to the net current,
conductivity/resistivity tensor inversion is not justified, because
the conductivities $\sigma_{xx}$ and $\sigma_{xy}$ are related to the
bulk of the 2D electron system. That is to say, a possible shunting
effect of the edge currents in the Hall bar (rectangular) geometry makes it
impossible to extract the value $\sigma_{xy}$ from the
magnetotransport data for $\rho_{xx}$ and $\rho_{xy}$.

\begin{figure}
\centering \scalebox{0.48}{\includegraphics[clip]{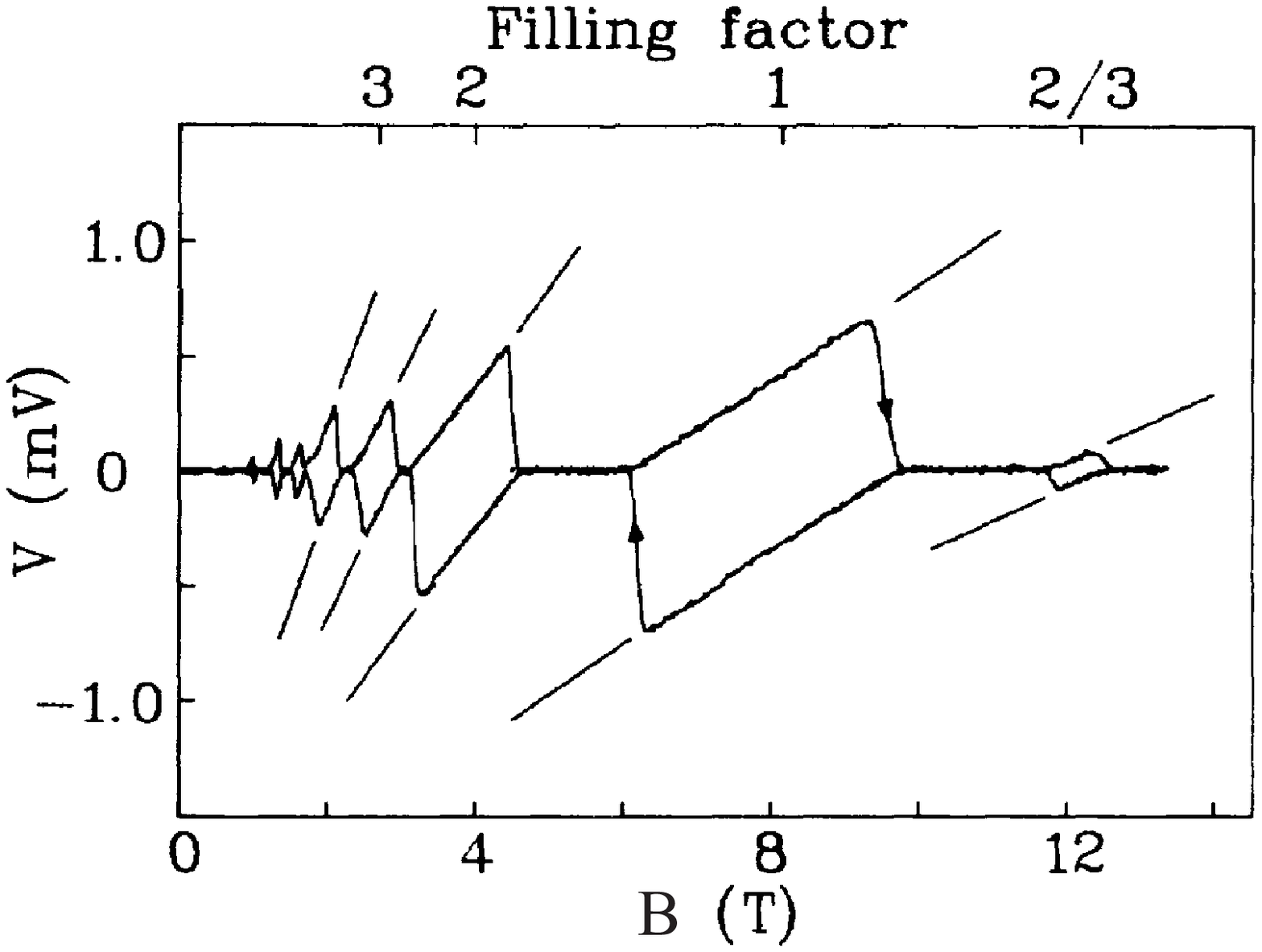}}
\caption{\label{SK:charge} The induced voltage in a Corbino sample of a
GaAs/AlGaAs heterostructure in up- and down-sweeps of the magnetic
field. Also shown by straight lines are the expected slopes for
$\nu=2/3$, 1, 2, 3, and 4. From Ref.~\cite{SK:dolgopolov92d}.}
\end{figure}

To verify whether or not the Hall conductivity is quantized, direct
measurements of $\sigma_{xy}$ are necessary, excluding the shunting
effect of the edge currents. Being equivalent to Laughlin's \textit{gedanken}
experiment~\cite{SK:laughlin81}, such measurements were realized using
the Corbino (ring) geometry which allows separation of the bulk contribution
to the net current (see, e.g., Ref.~\cite{SK:dolgopolov92d}). A Hall
charge transfer below the Fermi level between the borders of a Corbino
sample is induced by a magnetic field sweep through the generated
azimuthal electric field. If the dissipative conductivity
$\sigma_{xx}\rightarrow0$, no discharge occurs, allowing determination
of the transferred charge, $Q=\sigma_{xy}\pi
r_{\rm{eff}}^2c^{-1}\delta B$, where $r_{\rm{eff}}$ is the effective
radius. The induced voltage, $V=Q/C$, which is restricted due to a
large shunting capacitance, $C$, changes linearly with magnetic field
with a slope determined by $\sigma_{xy}$ in the quantum Hall states
until the dissipationless quantum Hall state breaks down
(Fig.~\ref{SK:charge}). The fact that the quantization accuracy of
$\sigma_{xy}$ (about 1\%) is worse compared to that of $\rho_{xy}$
may be attributed to non-constancy of the effective area in not very
homogeneous samples. Thus, the Hall current in the QHE flows not only at the edges but also in the bulk of the 2D
electron system through the extended states in the filled Landau
levels.

The finite Hall conductivity measured in the Corbino geometry in the
arrangement of Laughlin's \textit{gedanken} experiment establishes the
existence of extended states in the Landau levels for both strongly
and weakly interacting 2D electron systems. Note that the
insignificance of edge-channel effects in transport experiments is
verified in the usual way by coincidence of the results obtained in
Hall bar and Corbino geometries.

\section{Metal-insulator Transitions in Perpendicular Magnetic Fields}

Metal-insulator transitions were studied for the quantum Hall phases
and the insulating phase at low electron densities. The insulating
phase was attributed to possible formation of a pinned Wigner crystal
\cite{SK:andrei88,SK:pudalov90,SK:santos92a}. However, floating-up of the
extended states relative to the Landau level centers and a close
similarity of all insulating phases have been found experimentally
\cite{SK:shashkin93,SK:shashkin94a,SK:shashkin94b}. Thus, the experimental
results excluded the formation of a pinned Wigner crystal in
available samples, but supported the existence of a metallic state in
zero field. It was also found that the bandwidth of the extended
states in the Landau levels is finite, which is in contradiction to
scaling arguments. Strangely, the latter experimental result has not
attracted much of theorists' attention.

\subsection{Floating-up of Extended States}
\label{SK:floating-up}

\begin{figure}
\centering \scalebox{0.63}{\includegraphics[clip]{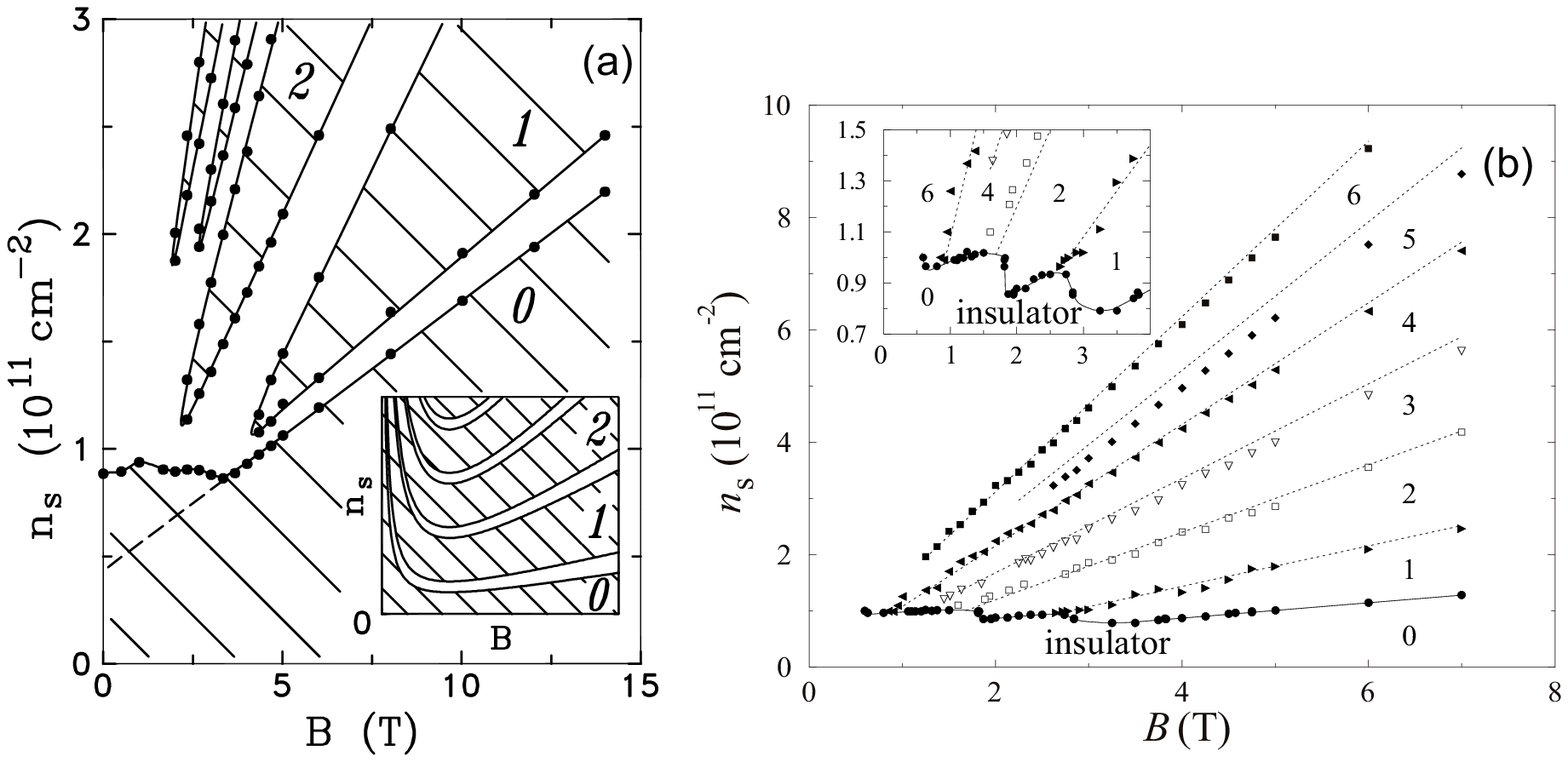}}
\caption{\label{SK:floating} (a)~Metal-insulator phase diagram in a
low-disordered 2D electron system in silicon MOSFETs obtained using a
cutoff criterion, $\sigma_{xx}=e^2/20h$, at a temperature $\approx
25$~mK. The dimensionless $\sigma_{xy}h/e^2$ in different insulating
phases is indicated. The slope of the dashed line is close to
$e/2hc$. A sketch of the expected phase diagram is displayed in the
inset. From Ref.~\cite{SK:shashkin93}. (b)~The map of extended states
determined by maxima in $\sigma_{xx}$ in a low-disordered silicon
MOSFET. Numbers show $\sigma_{xy}$ in units of $e^2/h$. From
Ref.~\cite{SK:kravchenko95b}.}
\end{figure}

The first experimental results on the metal-insulator phase diagram
at low temperatures in low-disordered silicon MOSFETs
\cite{SK:shashkin93} already revealed discrepancies with the theory
(Fig.~\ref{SK:floating}(a)). In that paper, a somewhat arbitrary
criterion for the longitudinal conductivity, $\sigma_{xx}=e^2/20h$,
was used to map out the phase boundary that corresponds to the
Anderson transition to the regime of strong localization. However,
first, the phase boundary was shown to be insensitive to the choice
of the cutoff value (see, e.g., Ref.~\cite{SK:dolgopolov92b}).  
Second, that particular cutoff value is consistent with the results
obtained for quantum Hall states by a vanishing activation energy
combined with a vanishing nonlinearity of current-voltage
characteristics when extrapolated from the insulating phase
\cite{SK:shashkin94a}.\footnote{Note that for the lowest-density phase boundary,
a lower value $\sigma_{xx}^{-1}\approx 100$~kOhm at a temperature
$\approx 25$~mK follows from the latter method.} The metallic phase
surrounds each insulating phase as characterized by the dimensionless
Hall conductivity, $\sigma_{xy}h/e^2$, that counts the number of
quantum levels below the Fermi level.\footnote{In bivalley (100)-silicon
MOSFETs, spin and valley degeneracies of the Landau level should be
taken into account.} This indicates that the extended states indeed
do not disappear discontinuously. Instead, with decreasing magnetic
field they float up in energy relative to the Landau level centers
and merge forming a metallic state in the limit of $B=0$ (for more on
this, see Sec.~\ref{SK:zero}). This contradicts the theoretical
scenario that in the limit of zero magnetic field the extended states
should float up indefinitely in energy~\cite{SK:khmelnitskii84} leading
to an insulating ground state. Besides, the experimental phase
boundary at low electron densities oscillates as a function of $B$
with minima corresponding to integer filling factors. The phase
boundary oscillations manifest themselves in that the magnetoresistance at electron densities near the $B=0$ metal-insulator transition oscillates with an amplitude that diverges as $T\rightarrow0$~\cite{SK:pudalov90}.  The regions in which the
magnetoresistance diverges are referred to as the reentrant insulating
phase.

\begin{figure}
\centering \scalebox{0.63}{\includegraphics[clip]{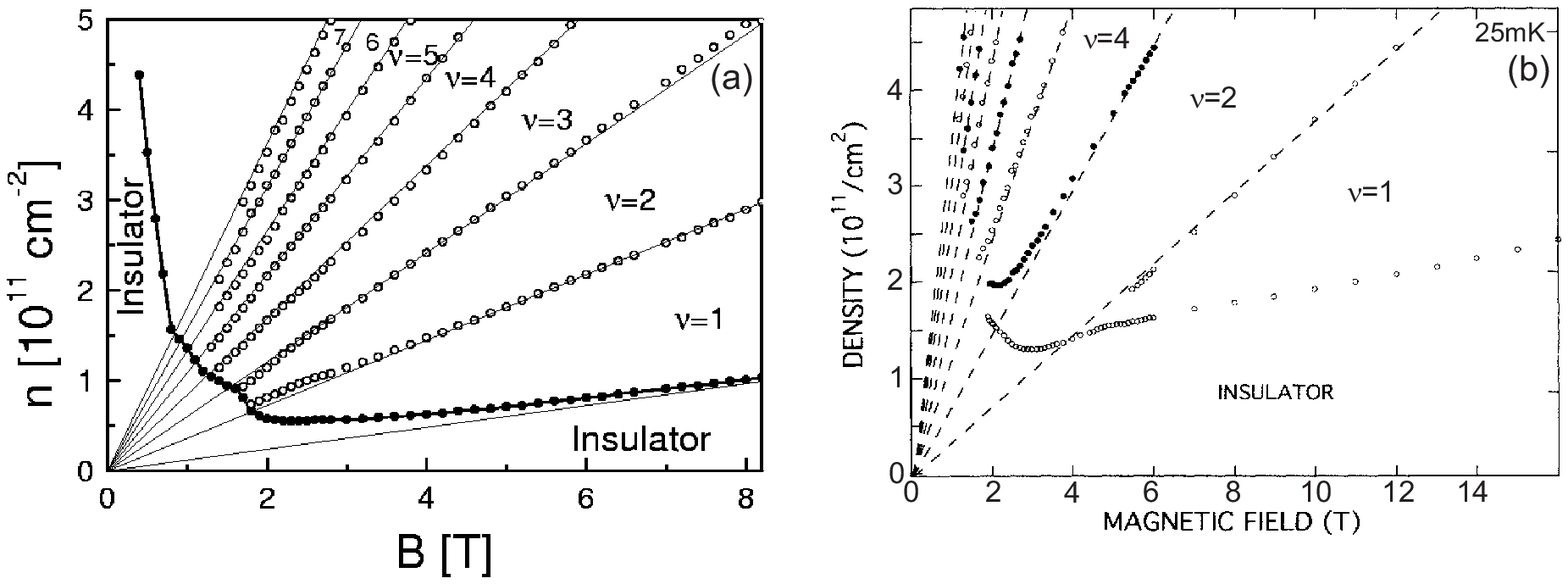}}
\caption{\label{SK:dultz} (a)~A map of the extended states for a
highly-disordered 2D hole system in a Ge/SiGe quantum well. The open
circles represent maxima in $\rho_{xx}$ and/or $d\rho_{xy}/dB$. The
solid circles correspond to crossing points of $\rho_{xx}$ at
different temperatures. Numbers show the value of $\sigma_{xy}h/e^2$.
Adapted from Ref.~\cite{SK:hilke00}. (b)~Behavior of the extended states
determined by maxima in $\sigma_{xx}$ in a strongly-disordered 2D
electron system in GaAs/AlGaAs heterostructures. Numbers show
$\sigma_{xy}$ in units of $e^2/h$. Adapted from
Ref.~\cite{SK:glozman95}.}
\end{figure}

The topology of the observed metal-insulator phase diagram\footnote{We refer here to merging of the extended states and, hence, the presence of direct
transitions between the insulating phase with $\sigma_{xy}=0$ and
quantum Hall phases with $\sigma_{xy}h/e^2>1$.} is robust, being
insensitive to the method for spotting the phase boundary
\cite{SK:shashkin94a,SK:kravchenko95b} and to the choice of 2D carrier
system~\cite{SK:hilke00}. This robustness was verified using a criterion of vanishing
activation energy and vanishing nonlinearity of current-voltage
characteristics as extrapolated from the insulating phase, 
allowing more accurate determination of the Anderson transition
\cite{SK:shashkin94a}. A method that had been suggested in
Ref.~\cite{SK:glozman95} was also applied for similar silicon MOSFETs
\cite{SK:kravchenko95b}. The extended states were studied by tracing
maxima in the longitudinal conductivity in the ($B,n_s$) plane
(Fig.~\ref{SK:floating}(b)) and good agreement with the aforementioned
results was found. A similar merging of at least the two lowest
extended states was observed in a more strongly disordered 2D hole system in a
Ge/SiGe quantum well~\cite{SK:hilke00} (Fig.~\ref{SK:dultz}(a)). The
extended states were associated either with maxima in $\rho_{xx}$
and/or $d\rho_{xy}/dB$, or with crossing points of $\rho_{xx}$ at
different temperatures. It is noteworthy that a bad combination of
the criterion for determining the phase boundary and the 2D carrier
system under study may lead to a failure in mapping out the phase
diagram down to relatively weak magnetic fields. In
Ref.~\cite{SK:glozman95}, extended states were studied by measuring
maxima in the longitudinal conductivity in the ($B,n_s$) plane for
the strongly-disordered 2D electron system in GaAs/AlGaAs
heterostructures (Fig.~\ref{SK:dultz}(b)). Because of strong damping of
the Shubnikov-de~Haas oscillations in low magnetic fields, the
desired region on the phase diagram below 2~T was not accessible in
that experiment. This invalidates the claim of Glozman \textit{et
al}.~\cite{SK:glozman95} that the extended states do not merge. The
behavior of the lowest extended state in Fig.~\ref{SK:dultz}(b), which
Glozman \textit{et al}.~\cite{SK:glozman95} claim to float up above the
Fermi level as $B\rightarrow0$, simply reflects the occurrence of a
phase boundary oscillation minimum at filling factor $\nu=2$, similar
to both the minimum at $\nu=1$ in Fig.~\ref{SK:dultz}(a) and to the case
of silicon MOSFETs (Fig.~\ref{SK:floating}). Such a minimum manifests
itself in that there exists a minimum in $\rho_{xx}$ at integer
$\nu\ge1$ that is straddled by the insulating phase. 

To this end, all
available data for the metal-insulator phase diagrams agree well with
each other, except those in the vicinity of $B=0$. In weak magnetic
fields, experimental results obtained in 2D electron systems with
high disorder are not method-independent. Glozman \textit{et al}.
\cite{SK:glozman95} found that the cutoff criterion yields basically a
flat phase boundary towards $B=0$, which is in agreement with the
data for silicon MOSFETs (Fig.~\ref{SK:floating}(a)). On the contrary,
Hilke et al.~\cite{SK:hilke00} employed the method based on temperature
dependencies of $\rho_{xx}$ and obtained a turn up on the phase
boundary in Fig.~\ref{SK:dultz}(a). Note that the validity of the data
for the lowest extended state at magnetic fields $\leq1.5$~T in
Fig.~\ref{SK:dultz}(a) is questionable because the weak temperature
dependencies of $\rho_{xx}$ as analyzed by Hilke \textit{et al}.
\cite{SK:hilke00} cannot be related to either an insulator or a metal.

As a matter of fact, the weak-field problem, whether or not there
is an indefinite rise of the phase boundary as $B\rightarrow0$, is
a problem of the existence of a metal-insulator transition at $B=0$
and $T=0$. In dilute 2D electron systems with low enough disorder,
the resistivity, $\rho$, strongly drops with decreasing temperature 
\cite{SK:sarachik99,SK:abrahams01}, providing an independent way of facing
the issue. Given strong temperature dependencies of $\rho$, those with
$d\rho/dT>0$ ($d\rho/dT<0$) can be associated with a metallic
(insulating) phase~\cite{SK:sarachik99,SK:abrahams01}. If extrapolation of
the temperature dependencies of $\rho$ to $T=0$ is valid, the curve
with $d\rho/dT=0$ should correspond to the metal-insulator transition
(see Sec.~\ref{SK:zero}). As long as in more-disordered 2D carrier
systems the metallic ($d\rho/dT>0$) behavior is suppressed (see,
e.g., Refs.~\cite{SK:hanein98,SK:pudalov01}) or disappears entirely, it is
definitely incorrect to extrapolate those weak temperature
dependencies of $\rho$ to $T=0$ with the aim to distinguish between
insulator and metal.

Another point at which one can compare experiment and theory is the oscillating
behavior of the phase boundary that restricts the insulating phase
with $\sigma_{xy}=0$ (see, e.g., Fig.~\ref{SK:floating}). Note that the
oscillations persist down to the magnetic fields corresponding to the
fillings of more than one Landau level. The oscillation period
includes the following stages. With decreasing magnetic field the
lowest extended states follow the Landau level, float up in energy
relative to its center, and merge with extended states in the next
quantum level. No merging was present in the original theoretical
considerations~\cite{SK:khmelnitskii84,SK:kivelson92,SK:khmelnitskii92},
leading to discrepancies between experiment and theory. Recently,
theoretical efforts have been concentrated on modifications of the
global phase diagram for the QHE to reach topological
compatibility with the observed metal-insulator phase diagram.
Although floating and/or merging of the extended states can be
obtained in the calculations, the oscillations of the phase boundary
at low electron densities have not yet been described theoretically.

\subsection{Similarity of the Insulating Phase and Quantum Hall Phases}
\label{SK:similarity}

The insulating phase at low electron densities was considered to be a
possible candidate for a pinned Wigner crystal. It was argued that its
aforementioned reentrant behavior is a consequence of the competition
between the QHE and the pinned Wigner crystal
\cite{SK:pudalov90}. Another supporting argument was strongly nonlinear
current-voltage characteristics in the insulating phase which were
attributed to depinning of the Wigner crystal. Similar features of
the insulating phase in a 2D electron (near $\nu=1/5$)
\cite{SK:andrei88} and 2D hole (near $\nu=1/3$)~\cite{SK:santos92a} systems
in GaAs/AlGaAs heterostructures with relatively low disorder were
also attributed to a pinned Wigner crystal which is interrupted by
the fractional quantum Hall state. An alternative scenario was
discussed in terms of percolation metal-insulator transition
\cite{SK:dolgopolov92b,SK:dolgopolov92a,SK:dolgopolov92c}. To distinguish
between the two scenarios, the behavior of activation energy and
current-voltage characteristics in the insulating phase was studied
and compared to that in quantum Hall phases
\cite{SK:shashkin94a,SK:shashkin94b}.

\begin{figure}
\centering \scalebox{0.62}{\includegraphics[clip]{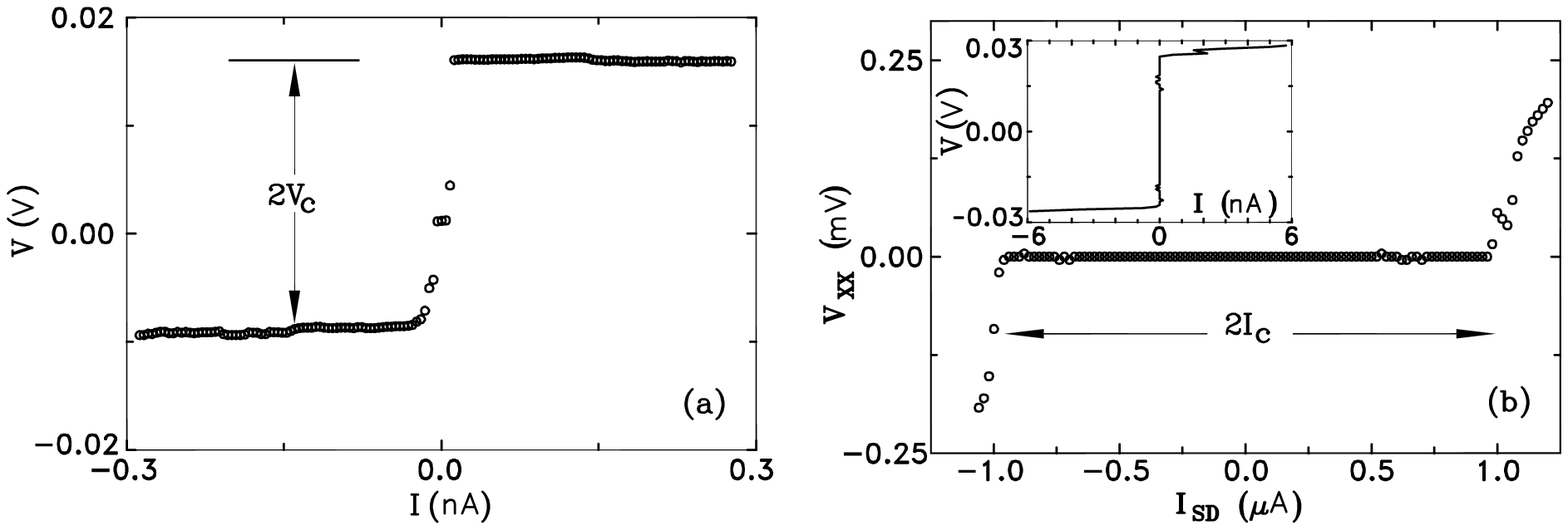}}
\caption{\label{SK:iv} Current-voltage characteristics in a
low-disordered silicon MOSFET in $B=12$~T at $T\approx 25$~mK for the
low-density insulating phase at $n_s=1.74\times 10^{11}$~cm$^{-2}$
(a) and the insulating phase with $\sigma_{xy}h/e^2=1$ at
$n_s=2.83\times 10^{11}$~cm$^{-2}$ (b). In (b) the measured breakdown
dependence $V_{xx}(I_{sd})$ is converted into current-voltage
characteristics (inset). From Ref.~\cite{SK:shashkin94a}.}
\end{figure}

\begin{figure}
\centering \scalebox{0.62}{\includegraphics[clip]{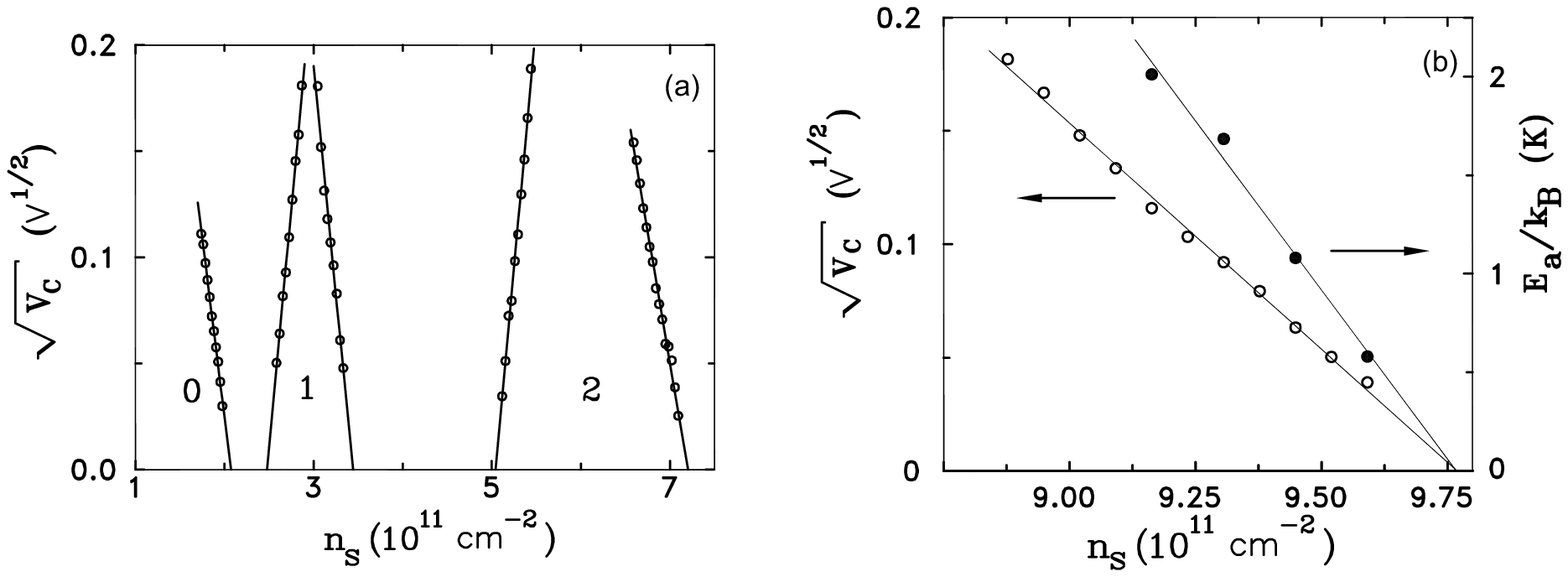}}
\caption{\label{SK:Vc} (a)~Square root of the critical voltage as a
function of electron density at the phase boundaries corresponding to
$\sigma_{xy}h/e^2=0$, 1, and 2 in $B=12$~T for a low-disordered 2D
electron system in silicon MOSFETs. (b)~Behavior of the critical
voltage and the activation energy near the phase boundary in
$B=16$~T. From Ref.~\cite{SK:shashkin94a}.}
\end{figure}

In contrast to the low-density insulating phase, the way of
determining the current-voltage characteristics of the quantum Hall
phases is different for Corbino and Hall bar geometries. In the
former the dissipationless Hall current does not contribute to the
dissipative current that is proportional to $\sigma_{xx}$, allowing
straightforward measurements of current-voltage curves for all
insulating phases. In the latter the two current channels are
connected through edge channels (see Sec.~\ref{SK:proof}), and
current-voltage characteristics correspond to quantum-Hall-effect
breakdown curves. The dissipative backscattering current, $I$, that
flows between opposite edge channels  is balanced by the Hall current 
in the filled Landau levels associated with the longitudinal voltage,
$V_{xx}$. As long as $\sigma_{xx}\ll\sigma_{xy}$, the quantized value
of $\sigma_{xy}$ is a factor that allows determination of
$I=\sigma_{xy}V_{xx}$ and the Hall voltage, $V=I_{sd}/\sigma_{xy}$,
from the experimental breakdown dependence of $V_{xx}$ on
source-drain current, $I_{sd}$. The dependence $V(I)$ is a
current-voltage characteristic, which is equivalent to the case of
Corbino geometry~\cite{SK:shashkin94a} (Fig.~\ref{SK:iv}). Not only are the
current-voltage curves similar for all insulating phases, but they
also behave identically near the metal-insulator phase boundaries
(Fig.~\ref{SK:Vc}(a)). The dependence of the critical voltage, $V_c$, on
the distance from the phase boundary is close to a parabolic law
\cite{SK:dolgopolov92b}. The phase boundary position determined by a
vanishing $V_c$ is practically coincident with that determined by a
vanishing activation energy, $E_a$, of electrons from the Fermi level
$E_F$ to the mobility edge, $E_c$ (Fig.~\ref{SK:Vc}(b)). The value $E_a$
is determined from the temperature dependence of the conduction in
the linear interval of current-voltage curves, which is activated at
not too low temperatures~\cite{SK:adkins76}; note that it transforms
into variable range hopping as $T\rightarrow0$ (see below). The
activation energy changes linearly with the distance from the phase
boundary, reflecting constancy of the thermodynamic density of states
near the transition point (see also Sec.~\ref{SK:zero}). The
threshold behavior of the current-voltage characteristics is caused
by the breakdown in the insulating phases. The breakdown occurs when
the localized electrons at the Fermi level gain enough energy to
reach the mobility edge in an electric field, $V_c/d$, over a
distance given by the localization length, $L$
\cite{SK:shashkin94a,SK:polyakov93}:
\begin{equation}eV_cL/d=|E_c-E_F|,\label{SK:break}\end{equation}
where $d$ is the corresponding sample dimension. The values $E_a$ and
$V_c$ are related through the localization length which is
temperature independent and diverges near the transition as
$L(E_F)\propto |E_c-E_F|^{-s}$ with exponent $s$ close to unity, in
agreement with the theoretical value $s=4/3$ in the classical percolation
problem~\cite{SK:shklovskii84}. The value of the localization length is
practically the same near all metal-insulator phase boundaries, which
indicates that even quantitatively, all insulating phases are very
similar. Note that since the localization length in Eq.~(\ref{SK:break})
is small compared to the sample sizes, the phase boundary position
determined by the diverging localization length refers to an infinite
2D system. As inferred from the vanishing of both $E_a$ and $V_c$ at
the same point (see Fig.~\ref{SK:Vc}(b)), possible shifts of the
mobility threshold due to finite sample dimensions are small, which
justifies extrapolations to the limit of $L\rightarrow\infty$.

The consequences of the method include the following.  (i)~As long as no
dramatic changes occur in transport properties, this excludes the pinned
Wigner solid as the origin for the insulating phase at low electron
densities in available samples of low-disordered silicon MOSFETs.
(ii)~The metal-insulator phase diagram of Fig.~\ref{SK:floating}(a) is
verified and substantiated. (iii)~The existence of a metal-insulator
transition in zero magnetic field is supported (see
Sec.~\ref{SK:zero}). (iv)~The bandwidth of the extended states in
the Landau levels is finite. All of these are also valid for
relatively low-disordered 2D carrier systems in GaAs/AlGaAs
heterostructures with the distinction that fractional quantum Hall
phases are involved. Yet, the topology of the phase diagram remains
unchanged, including the oscillating behavior of the phase boundary
that restricts the low-density insulating phase. Additional
confirmation of the percolation transition to the low-density
insulating phase in GaAs/AlGaAs heterostructures was obtained by
studies of the high-frequency conductivity~\cite{SK:li94} and
time-resolved photoluminescence of 2D electrons~\cite{SK:kukushkin93},
as discussed in Ref.~\cite{SK:shashkin94b}.

The insulating phase at low electron densities is special in what
follows. Deep in the insulating state and at low temperatures the
variable-range-hopping regime  occurs in which the conductivity
$\sigma_{xx}$ is small compared to its peak value
\cite{SK:shklovskii84}. In this regime it was predicted that the
deviation, $\Delta\sigma_{xy}$, of $\sigma_{xy}$ from its quantized
value in strong magnetic fields is much smaller than
$\sigma_{xx}\propto\exp(-(T_0/T)^{1/2})$~\cite{SK:wysokinski83}:
$\Delta\sigma_{xy}\propto\sigma_{xx}^\gamma$ with exponent
$\gamma\approx 1.5$. A finite $\rho_{xy}$ contrasted by diverging
$\rho_{xx}$ was found in calculations of the $T=0$ magnetotransport
coefficients in the insulating phase with vanishing $\sigma_{xx}$ and
$\sigma_{xy}$~\cite{SK:viehweger90}. Such a behavior of $\rho_{xx}$ and
$\rho_{xy}$ indicates a special quadratic relation between
conductivities: $\sigma_{xy}\propto\sigma_{xx}^2$. Moreover, it was
shown that $\rho_{xy}$ is close to the classical value ($B/n_sec$)
\cite{SK:zhang92}, providing arguments for the existence of a Hall
insulator phase~\cite{SK:kivelson92}. Indeed, values $\rho_{xy}$ close
to $B/n_sec$ were experimentally found in the low-density insulating
phase. Thus, the distinction of the Hall insulator phase from the
quantum Hall phases, i.e., the absence of extended states below the
Fermi level, becomes evident when expressed in terms of
$\rho_{xx}$ and $\rho_{xy}$.

\subsection{Scaling and Thermal Broadening}
\label{SK:scaling}

It was predicted that the localization length diverges as a power law
at a single energy, $E^*$, which is the center of the Landau level
\cite{SK:iordansky82}: $L(E)\propto |E-E^*|^{-s}$. An idea to check this
prediction based on low-temperature measurements of $\sigma_{xx}$
\cite{SK:aoki85} was quickly developed to a concept of single-parameter
scaling~\cite{SK:pruisken88}. It was suggested that the
magnetoresistance tensor components are functions of a single
variable that is determined by the ratio of the dephasing length,
$L_d(T)\propto T^{-p/2}$ (where $p$ is the inelastic-scattering-time
exponent), and the localization length. The concept was claimed to be
confirmed by measurements of temperature dependencies of the peak
width, $\Delta B$, in $\rho_{xx}$ (or $\sigma_{xx}$) and the maximum
of $d\rho_{xy}/dB$ in a highly-disordered 2D electron system in
InGaAs/InP heterostructures, yielding $\Delta B\propto
T^\kappa$, where $\kappa=p/2s\approx 0.4$~\cite{SK:wei88}. Later, both
deviations in the power law and different exponents in the range
between $\kappa=0.15$ and $\kappa=1$ were observed for other 2D
carrier systems, different Landau levels, and different disorder
strengths (see, e.g.,
Refs.~\cite{SK:dolgopolov91a,SK:koch91a,SK:wakabayashi89,SK:wei92}). Importantly,
the scaling analysis of experimental data in question is based on two
unverified assumptions: (i)~zero bandwidth of the extended states in
the Landau levels; and (ii)~constancy of the thermodynamic density of
states in the scaling range. If either assumption is not valid, this
may lead, at least, to underestimating the experimental value of
exponent $\kappa$.

\begin{figure}
\centering \scalebox{0.61}{\includegraphics[clip]{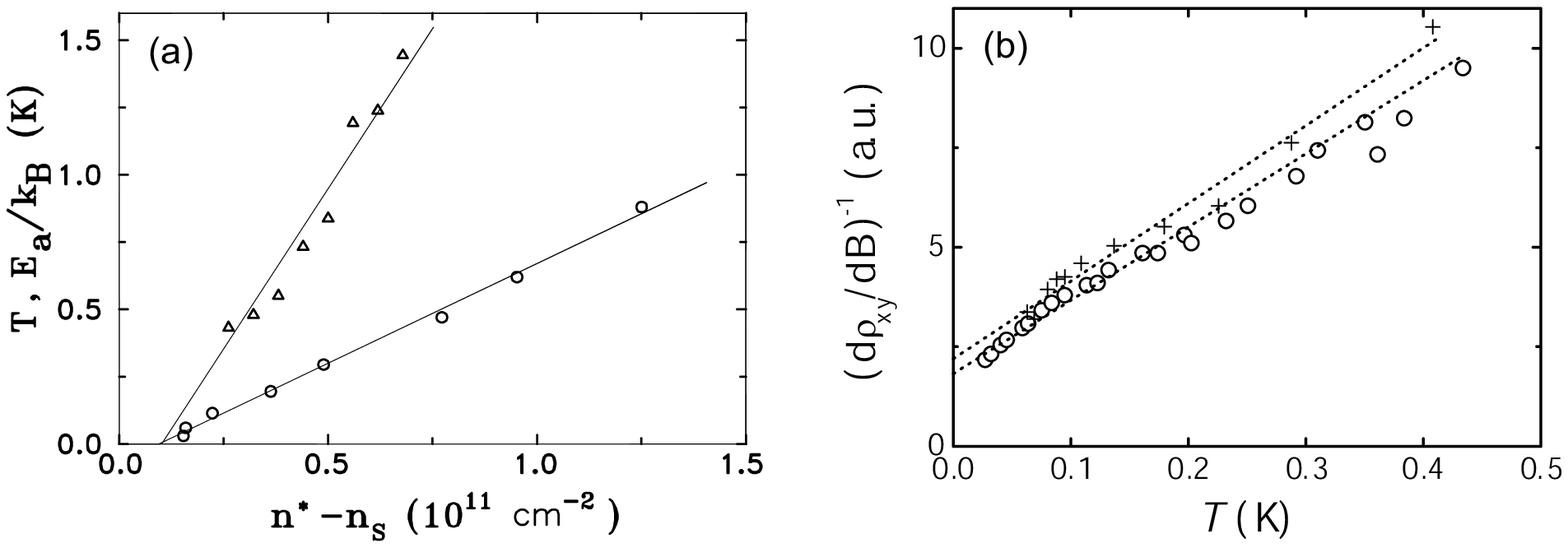}}
\caption{\label{SK:width} (a)~Temperature dependence of the $\rho_{xx}$
peak width ($n^*-n_s$) at half of the peak height counted from $n^*$
corresponding to $\nu^*=2.5$ (circles) and the behavior of the
activation energy (triangles) in a low-disordered silicon MOSFET in
$B=14$~T. From Ref.~\cite{SK:shashkin94a}. (b)~Temperature dependence of
the $\rho_{xx}$ peak width as determined by the maximum
$d\rho_{xy}/dB$ at $\nu^*=1.5$ in a highly-disordered 2D electron
system in GaAs/AlGaAs heterostructures. Different symbols correspond
to different runs. The dashed lines are linear fits to the data.
Adapted from Ref.~\cite{SK:wei92}.}
\end{figure}

The method of vanishing activation energy and vanishing nonlinearity
of current-voltage characteristics as extrapolated from the
insulating phase shows that the former assumption is not justified.
Also, measurements of the peak width in $\rho_{xx}$ as a function of
temperature in low-disordered silicon MOSFETs yield a linear
dependence which extrapolates to a finite peak
width~\cite{SK:shashkin94a} as $T\rightarrow0$ (Fig.~\ref{SK:width}(a)). Very
similar temperature and frequency dependencies were observed in
highly-disordered 2D carrier systems in GaAs/AlGaAs heterostructures
\cite{SK:balaban98} and Ge/SiGe heterostructures~\cite{SK:hilke97}. It is
noteworthy that a similar behavior is revealed if the data from the
publications, which claim the observation of scaling, is plotted on a
linear rather than logarithmic scale (see, e.g.,
Fig.~\ref{SK:width}(b)); finite values of the peak width as
$T\rightarrow0$ are even more conspicuous for the data of
Refs.~\cite{SK:dolgopolov91a,SK:koch91a,SK:li09}. The reason for the ambiguity
is quite simple: within experimental uncertainty, it is difficult, especially on a logarithmic scale, to distinguish between
sublinear/superlinear fits to the data and linear fits which do not
have to run through the origin. Note that attempts were made to
relate the finite peak width as $T\rightarrow0$ to the dephasing
length reaching the sample size~\cite{SK:koch91a,SK:li09}. However, the
suggested finite-size effect is not supported by experimental data,
because in different samples with different sizes, the disorder is
also different.  It is the disorder, rather than the sample size,
that may be responsible for the behavior of the values measured in
different samples.

Although lack of data in most of the above experimental papers does
not allow one to verify the validity of both assumptions, it is very
likely that there is no qualitative difference between all of the
discussed results. As a matter of fact, they can be described by a
linear, or weakly sub-linear temperature dependence with a finite
offset at $T=0$. This is concurrent with the results obtained by
vanishing activation energy and vanishing nonlinearity of
current-voltage characteristics as extrapolated from the insulating
phase. So, the single-parameter scaling is not confirmed by the
experimental data which establish the finite bandwidth of the
extended states in the Landau levels. 

There is an alternative and simple explanation of the temperature
dependence of the peak width in $\rho_{xx}$ in terms of thermal
broadening. Within a percolation picture, if the activation energy
$E_a\sim k_BT$, the conduction is of the order of the maximum
$\sigma_{xx}$ so that the value of $\sim k_BT$ gives a thermal shift
of the effective mobility edge corresponding to the $\sigma_{xx}$
peak width~\cite{SK:shashkin94a}. Although the concept of thermal
broadening has been basically ignored in the literature in the search for
less trivial data interpretations, it looks as if no experimental
results go beyond this, favoring the concept of single-parameter
scaling. Once the behavior of the localization length is not
reflected by the temperature-dependent peak width in $\rho_{xx}$, no
experimental support is provided for numerical calculations of the
localization length which give a somewhat larger exponent $s\approx
2$ compared to $s=4/3$ in classical percolation problem (see, e.g.,
Ref.~\cite{SK:huckestein95}).

\section{Zero-field Metal-insulator Transition}
\label{SK:zero}

In contrast to the case of quantizing magnetic fields, no extended
states are expected in zero magnetic field, at least for
weakly-interacting 2D electron systems. The criterion of vanishing
activation energy and vanishing nonlinearity of current-voltage
characteristics as extrapolated from the insulating phase, however,
results in an opposite conclusion.  To sort out this inconsistency, further support by independent
experimental verifications is needed.

\begin{figure}
\centering \scalebox{0.62}{\includegraphics[clip]{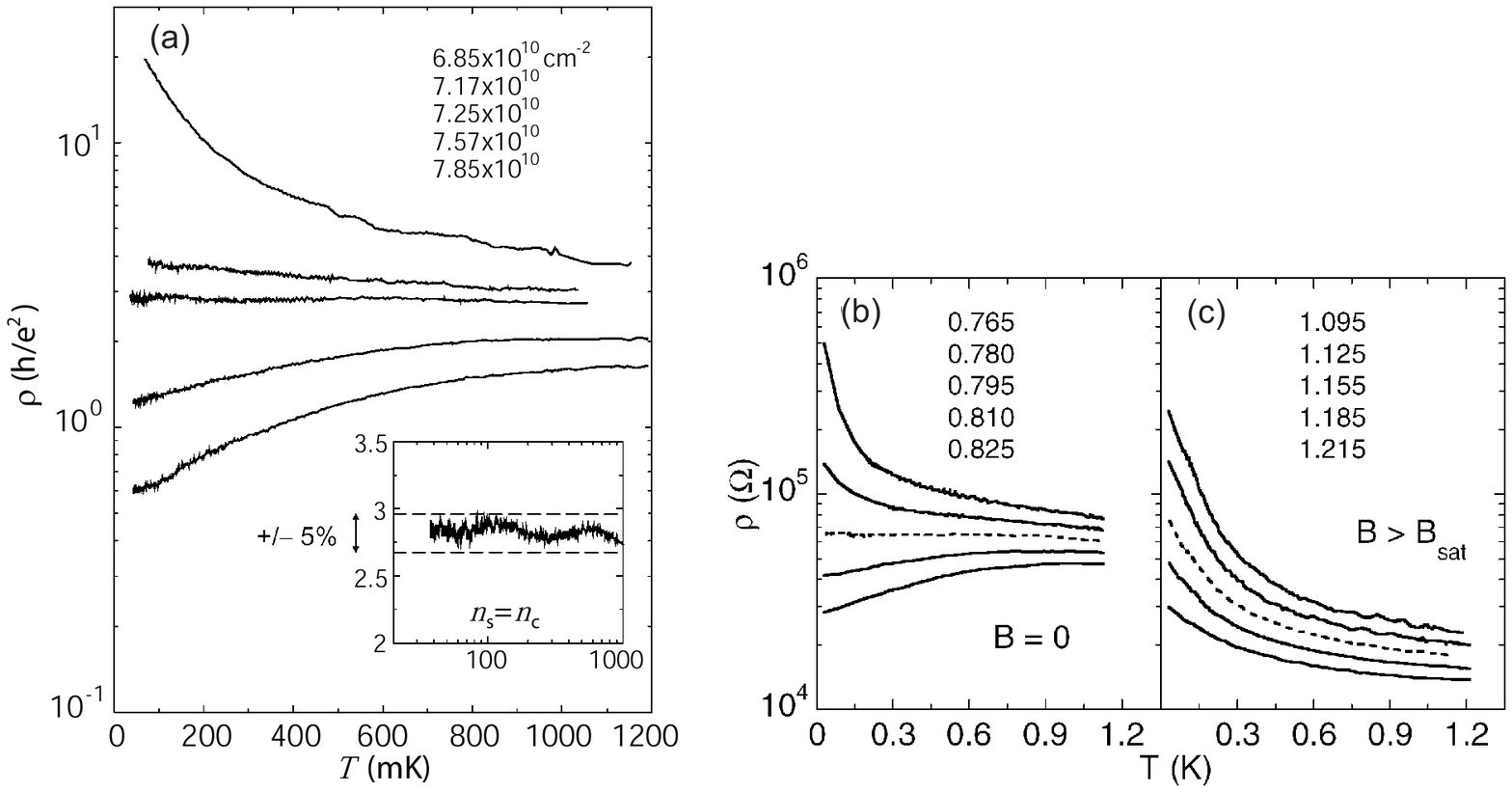}}
\caption{\label{SK:flat} (a)~Resistivity as a function of temperature at
different electron densities in a low-disordered silicon MOSFET. The
inset shows the middle curve on an expanded scale. From
Ref.~\cite{SK:kravchenko00b}. (b, c)~Temperature dependence of the
resistivity of a low-disordered silicon MOSFET at different electron
densities near the metal-insulator transition, (b) in zero magnetic field
and (c) in a parallel magnetic field of 4~T. The electron
densities are indicated in units of $10^{11}$~cm$^{-2}$. From
Ref.~\cite{SK:shashkin01b}.}
\end{figure}

\begin{figure}
\centering \scalebox{0.6}{\includegraphics[clip]{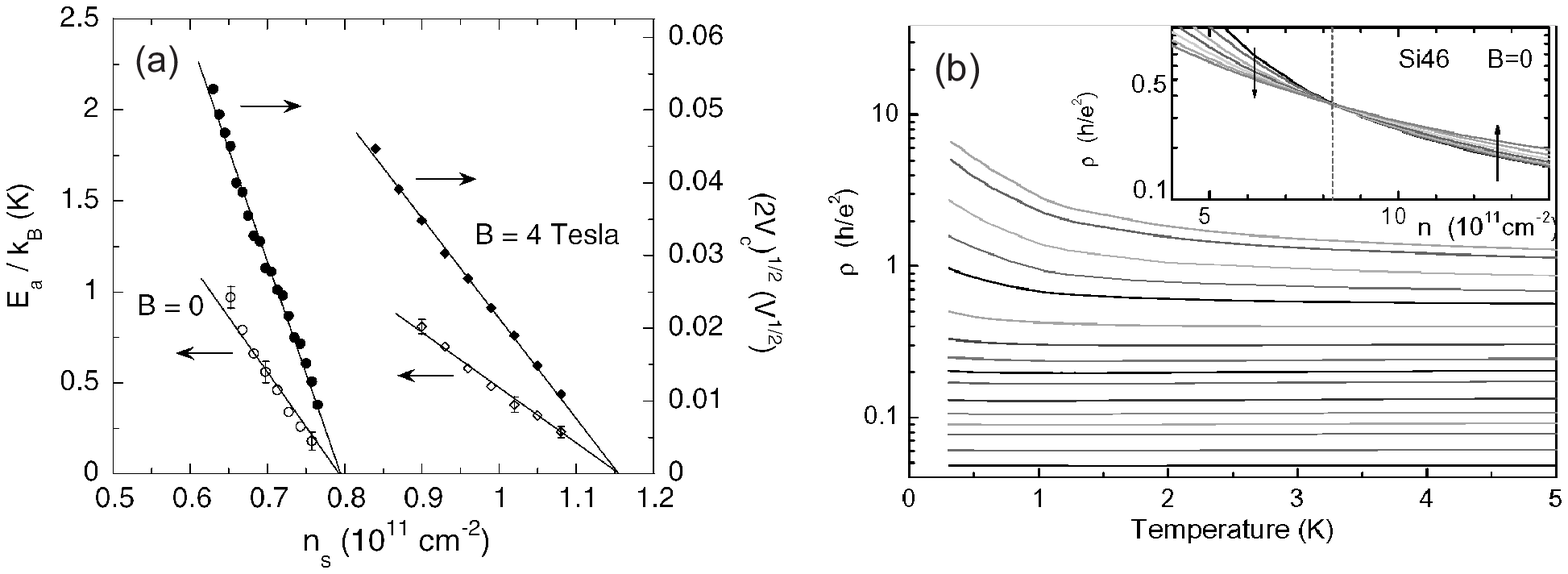}}
\caption{\label{SK:nc2} (a)~Activation energy and square root of the
threshold voltage as a function of electron density in zero magnetic
field (circles) and in a parallel magnetic field of 4~T (diamonds)
for the same silicon MOSFET as in Fig.~\ref{SK:flat}(b, c). The critical
densities correspond to the dashed lines in Fig.~\ref{SK:flat}(b,c).
From Ref.~\cite{SK:shashkin01b}. (b)~Resistivity versus temperature in a
strongly-disordered silicon MOSFET at the following electron
densities: 3.85, 4.13, 4.83, 5.53, 6.23, 7.63, 9.03, 10.4, 11.8,
13.2, 16.0, 18.8, 21.6, 24.4, 30.0, and $37.0\times
10^{11}$~cm$^{-2}$. The $\rho(n_s)$ isotherms are shown in the inset.
Adapted from Ref.~\cite{SK:pudalov01}.}
\end{figure}

Another criterion is based on the analysis of the temperature
dependencies of the resistivity at $B=0$. Provided these are strong,
those with positive (negative) derivative $d\rho/dT$ are indicative
of a metal (insulator)~\cite{SK:sarachik99,SK:abrahams01}; note that in the
vicinity of the transition, $\rho(T)$ dependencies obey the scaling
law with exponent $\kappa\approx1$, which is consistent with the
concept of thermal broadening/shift by the value $\sim k_BT$ of the
effective mobility edge in the insulating phase (see
Sec.~\ref{SK:scaling}). If extrapolation of $\rho(T)$ to $T=0$ is
valid, the critical point for the metal-insulator transition is given
by $d\rho/dT=0$. In a low-disordered 2D electron system in silicon
MOSFETs, the resistivity at a certain electron density shows
virtually no temperature dependence over a wide range of temperatures
\cite{SK:sarachik99,SK:kravchenko00b} (Fig.~\ref{SK:flat}(a)). This curve
separates those with positive and negative $d\rho/dT$ nearly
symmetrically at temperatures above 0.2~K~\cite{SK:abrahams01}. Assuming
that it remains flat down to $T=0$, one obtains the critical point
which corresponds to a resistivity $\rho\approx 3h/e^2$.

Recently, these two criteria have been applied simultaneously to the
2D metal-insulator transition in low-disordered silicon MOSFETs
\cite{SK:shashkin01b,SK:jaroszynski02}. In zero magnetic field, both
methods yield the same critical density $n_c$ (Figs.~\ref{SK:flat}(b)
and \ref{SK:nc2}(a)). Since one of the methods is temperature
independent, this equivalence strongly supports the existence of a
metal-insulator transition at $T=0$ in $B=0$. This also adds
confidence that the curve with zero derivative $d\rho/dT$ will remain
flat (or at least will retain finite resistivity value) down to zero
temperature. Additional confirmation in favor of zero-temperature
zero-field metal-insulator transition is provided by magnetic
measurements~\cite{SK:shashkin05}, as described in the next section. It
is argued that the metal-insulator transition in silicon
samples with very low disorder potential is driven by interactions.
This is qualitatively different from a localization-driven transition
in more-disordered samples that occurs at appreciably higher
densities.

For 2D electron systems both with high disorder in zero magnetic
field (see Sec.~\ref{SK:floating-up}) and in parallel magnetic
fields, the metallic ($d\rho/dT>0$) behavior is suppressed
\cite{SK:hanein98,SK:pudalov01,SK:shashkin01b,SK:shashkin06a} or disappears
entirely, and extrapolation of the weak $\rho(T)$ dependence to
$T=0$ is not justified, invalidating the derivative criterion for the
critical point for the metal-insulator transition
(Figs.~\ref{SK:flat}(c) and \ref{SK:nc2}(b)). Once one of the two methods
fails, it remains to be seen how to verify the conclusion as inferred
from the other method. This makes uncertain the existence of a
zero-temperature metal-insulator transition in 2D electron systems
both with high disorder in zero magnetic field and in parallel
magnetic fields.

Owing to its simplicity, the derivative
method is widely used for describing metallic ($d\rho/dT>0$) and
insulating ($d\rho/dT<0$) temperature dependencies of resistance in a
restricted temperature range. However, to avoid confusion with metallic and
insulating phases, one should employ alternative methods for
determining the metal-insulator transition point. Such methods,
including a vanishing activation energy and noise measurements, have
been applied to highly-disordered 2D carrier systems
\cite{SK:jaroszynski02,SK:bogdanovich02}. Being similar, they yield lower
critical densities $n_c$ for the metal-insulator transition compared
to those obtained using formally the derivative criterion. This
simply reflects the fact that the metallic ($d\rho/dT>0$) behavior is
suppressed. The critical density $n_c$, at which the exponential
divergence of the resistivity as $T\rightarrow0$ ends, increases
naturally with disorder strength. It also increases somewhat with
parallel magnetic field, saturating above a certain field, as was
found in dilute silicon MOSFETs~\cite{SK:dolgopolov92c,SK:shashkin01b}.

\section{Possible Ferromagnetic Transition}

\begin{figure}
\centering \scalebox{0.62}{\includegraphics[clip]{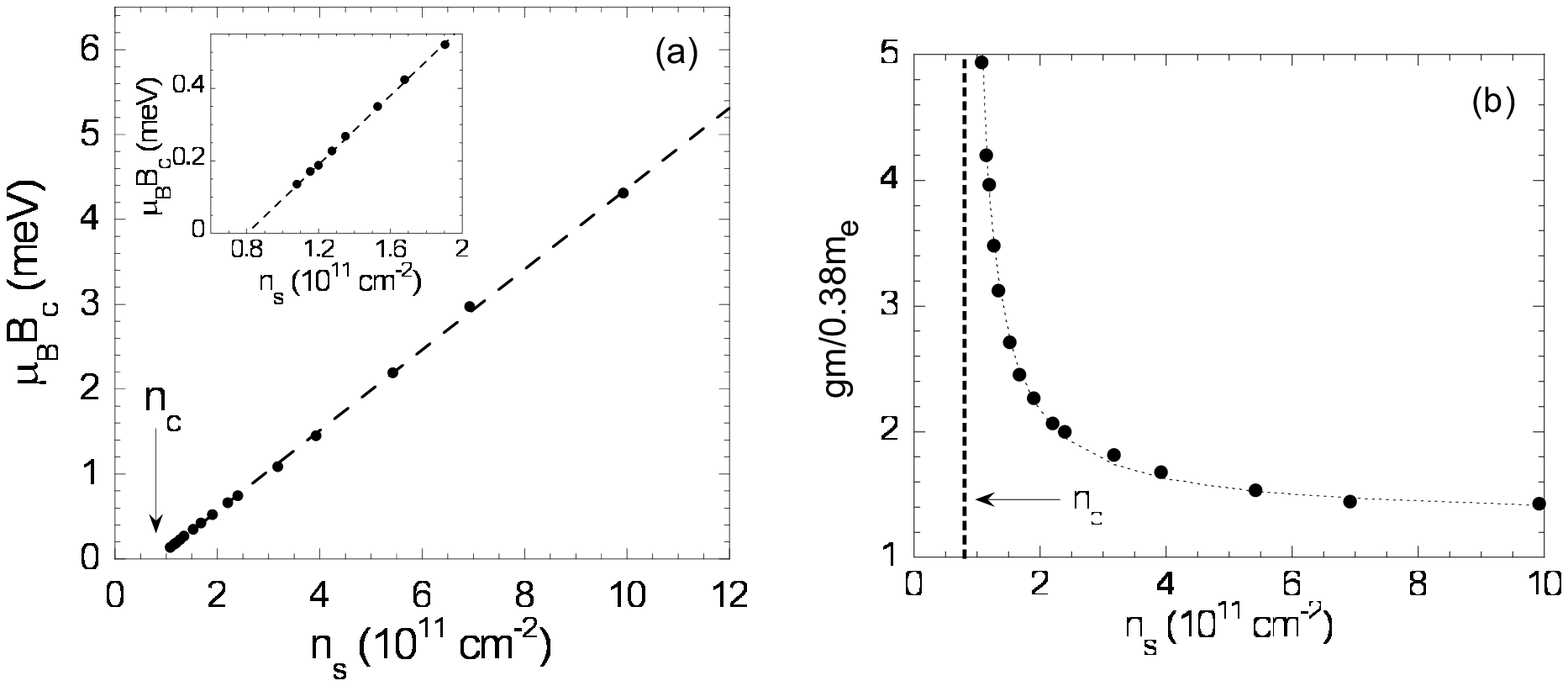}}
\caption{\label{SK:Bc} (a)~Dependence of the polarization field on
electron density, obtained by scaling the parallel-field
magnetoresistance in the spirit of Ref.~\cite{SK:dolgopolov00}, in a low-disordered silicon MOSFET. The dashed line
is a linear fit. The critical density $n_c$ is indicated. (b)~The
product $gm$ versus electron density obtained from the data for
$B_c$. From Ref.~\cite{SK:shashkin01a}.}
\end{figure}

\begin{figure}[t]
\centering \scalebox{0.64}{\includegraphics[clip]{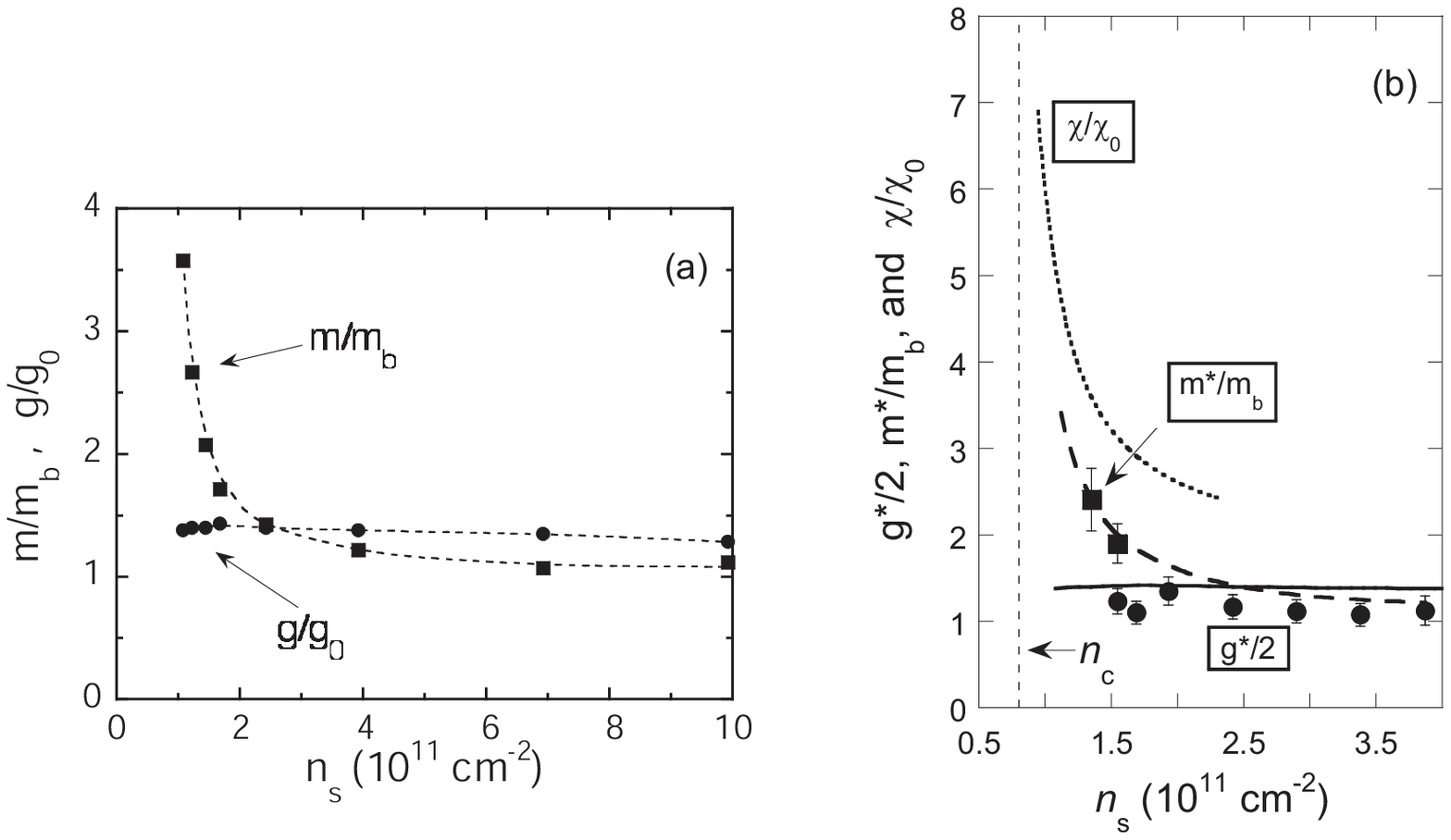}}
\caption{\label{SK:sigma} (a)~The effective mass and $g$ factor versus
electron density determined from an analysis of the
temperature-dependent conductivity~\cite{SK:zala01} and parallel-field
magnetoresistance. The dashed lines are guides to the eye. From
Ref.~\cite{SK:shashkin02b}. (b)~The effective mass (squares) and $g$
factor (circles), determined by magnetization measurements in perpendicular magnetic fields, as a function of the electron density. The solid and
long-dashed lines represent, respectively, the $g$ factor and
effective mass, previously obtained from transport measurements
\cite{SK:shashkin02b}, and the dotted line is the Pauli spin
susceptibility obtained by magnetization measurements in parallel
magnetic fields~\cite{SK:shashkin06b}. The critical density $n_c$ for
the metal-insulator transition is indicated. From
Ref.~\cite{SK:anissimova06}.}
\end{figure}

After a strongly enhanced ratio $gm$ of the spin and the cyclotron
splittings was found at low electron densities in silicon MOSFETs
\cite{SK:kravchenko00a}, it became clear that the system behavior was well
beyond the weakly interacting Fermi liquid. It was reported that the
parallel magnetic field required to produce complete spin
polarization, $B_c\propto n_s/gm$, tends to vanish at a finite
electron density $n_\chi\approx 8\times 10^{10}$~cm$^{-2}$, which is
close to the critical density $n_c$ for the metal-insulator
transition in this electron system~\cite{SK:shashkin01a,SK:vitkalov01,SK:kravchenko02}
(Fig.~\ref{SK:Bc}). These findings point to a sharp increase of the spin
susceptibility, $\chi\propto gm$, and possible ferromagnetic
instability in dilute silicon MOSFETs. The fact that $n_\chi$ is
close to the critical density $n_c$ indicates that the
metal-insulator transition in silicon samples with very low disorder
potential is a property of a clean 2D system and is driven by
interactions~\cite{SK:shashkin01a}. A similar although less pronounced
behavior was observed in other 2D carrier systems~\cite{SK:gao02}. The
experimental results indicated that in silicon MOSFETs it is the
effective mass, rather than the $g$ factor, that sharply increases at
low electron densities~\cite{SK:shashkin02b} (Fig.~\ref{SK:sigma}(a)). They
also indicated that the anomalous rise of the resistivity with
temperature is related to the increased mass. The magnitude of the
mass does not depend on the degree of spin polarization, which points
to a spin-independent origin of the effective mass enhancement
\cite{SK:shashkin03a}. It was found that the relative mass enhancement
is system- and disorder-independent and is determined by
electron-electron interactions only~\cite{SK:shashkin07}.

In addition to transport measurements, thermodynamic measurements of
the magnetocapacitance and magnetization of a 2D electron system in
low-disordered silicon MOSFETs were performed, and very similar
results for the spin susceptibility, effective mass, and $g$ factor
were obtained~\cite{SK:khrapai03a,SK:shashkin06b,SK:anissimova06}
(Fig.~\ref{SK:sigma}(b)). The Pauli spin susceptibility behaves
critically close to the critical density $n_c$ for the $B=0$
metal-insulator transition: $\chi\propto n_s/(n_s-n_\chi)$. This is
in favor of the occurrence of a spontaneous spin polarization (either
Wigner crystal or ferromagnetic liquid) at low $n_s$, although in
currently available samples, the residual disorder conceals the
origin of the low-density phase. The effective mass increases sharply
with decreasing density while the enhancement of the $g$ factor is
weak and practically independent of $n_s$. Unlike in the Stoner
scenario, it is the effective mass that is responsible for the
dramatically enhanced spin susceptibility at low electron densities.

Thus, the experimental results obtained in low-disordered silicon
MOSFETs indicate that on the metallic side the metal-insulator
transition is driven by interactions, while on the insulating side
this is still a classical percolation transition with no dramatic
effects from interactions. One can consider the metal-insulator
transition in the cleanest of currently available samples as a
quantum phase transition, even though the problem of the competition
between metal-insulator and ferromagnetic transitions is not yet
resolved. It is not yet clear whether or not electron crystallization
expected in the low-density limit is preceded by an intermediate
phase like ferromagnetic liquid.

\section{Outlook}

Critical analysis of the available experimental data for 2D electron
systems both in zero and in quantizing magnetic fields shows that
consequences of the scaling theory of localization for 
noninteracting 2D electrons are not confirmed. The main points to be
addressed by theory are the problem of finite bandwidth of the
extended states in the Landau levels and that of a quantum phase
transition in low-disordered 2D electron systems in zero magnetic
field, including the competition between metal-insulator and
ferromagnetic transitions. Recently, some progress has been made in
describing the behavior of low-disordered strongly interacting 2D
electron systems in zero magnetic field: it has been shown that the
metallic ground state can be stabilized by electron-electron
interactions~\cite{SK:punnoose05}. It is possible that it may also be
necessary to take into account electron-electron interactions to
describe the quantum phase transitions that are characterized by the
finite bandwidth of the extended states in the Landau levels.

The finding that in dilute 2D electron systems the spin
susceptibility tends to diverge due to strong increase in the
effective mass remains basically unexplained, and the particular
mechanism leading to the effect remains to be seen. It is worth
discussing the latest theoretical developments which are claimed to
be valid for the strongly-interacting limit. According to the
renormalization group analysis for multi-valley 2D systems, the
effective mass dramatically increases at disorder-dependent density
for the metal-insulator transition while the $g$ factor remains
nearly intact~\cite{SK:punnoose05}. However, the prediction of
disorder-dependent effective mass is in contradiction to the
experiment. Besides, the results of Ref.~\cite{SK:punnoose05} are valid
only in the near vicinity of the metal-insulator transition, while
the tendency of the spin susceptibility to diverge can be traced up
to the densities exceeding $n_c$ by a factor of a few. In the
Fermi-liquid-based model of Ref.~\cite{SK:khodel05}, a flattening at the
Fermi energy in the spectrum has been predicted that leads to a
diverging effective mass. Still, the expected dependence of the
effective mass on temperature is not confirmed by the experimental
data. The strong increase of the effective mass has been obtained, in
the absence of the disorder, by solving an extended Hubbard model
using dynamical mean-field theory~\cite{SK:pankov08}. This is consistent
with the experiment, especially taking into account that the relative
mass enhancement has been experimentally found to be independent of
the level of the disorder. The dominant increase of $m$ near the
onset of Wigner crystallization follows also from an alternative
description of the strongly-interacting electron system beyond the
Fermi liquid approach (see, \textit{e.g.}, Ref.~\cite{SK:spivak04}).

On the experimental side, progress in the fabrication of
increasingly high mobility Si, Si/SiGe, and GaAs-based devices will
open up the possibility of probing the intrinsic properties of clean 2D
electron systems at still lower densities, where electron-electron
interactions are yet stronger and, presumably, the previously
observed behaviors will be yet more pronounced. Moreover, as high-mobility devices made with other semiconductors become available,
further tests of the universality of the observed phenomena will add
to our knowledge of 2D quantum phase transitions.

\clearpage
\printindex

\end{document}